\pdfoutput=1
\documentclass[aps,prb,twocolumn,nofootinbib]{revtex4-2}
\usepackage{graphicx}
\usepackage{capt-of}
\usepackage{amsmath}
\usepackage{newtxtext,newtxmath}
\usepackage[dvipsnames]{xcolor}
\usepackage[colorlinks=true,linkcolor=MidnightBlue,urlcolor=MidnightBlue,citecolor=MidnightBlue]{hyperref}
\graphicspath{{figures/}}
\begin{document}

\title{QMBL: the Quantum Many-Body Leaderboard\\[3pt]A live record of published ground-state energies for lattice Hamiltonians and the electron gas}

\author{Tristan Zerweck}
\author{Juan Carrasquilla}
\affiliation{Institute for Theoretical Physics, ETH Zurich, 8093 Zurich, Switzerland}

\begin{abstract}
\noindent In condensed matter physics, scientists explore phase diagrams of matter through various methods (analysis, numerics, experiment) and achieve consensus when enough methods agree~\cite{LeBlanc2015}. The quality of variational results for quantum many-body ground states is mainly judged by energy values and respective errors. Here, we present QMBL, the Quantum Many-Body Leaderboard (\url{https://www.qmbl.org}), serving as the energy layer of the scientific consensus. It is a maintained, interactive record of published ground-state energies for lattice Hamiltonians and the homogeneous electron gas. The latest release v0.5.0 holds 2511 energies on 532 instances and collects all of the community's efforts into one view. Of those instances 285 have no exact solution (164 lattice and all 121 electron-gas instances) and any new record serves as a direct improvement to an open science problem. They are the targets of a companion benchmark, QMBL: Open Problems.
\end{abstract}

\maketitle

In 2024, Wu et al.\ standardized scientific results in the condensed matter physics community by introducing the V-score for measuring ground-state accuracy~\cite{Wu2024}. The accompanying dataset, VarBench, has not changed since October 2024 and inspired us to offer a live alternative. This leaderboard aggregates scattered results of ground-state energies with fragmented conventions (e.g.\ Pauli versus spin operators and energy per site versus total energies). Aggregation work that would have been tedious manual labour 4 years ago has been made cheap by AI agents, who can read every paper, convert conventions and check each number against its source (cf.\ Methods~\cite{qmblMethods}).

The physical models in this dataset include six lattice Hamiltonians, namely Heisenberg, Hubbard, Impurity, $J_1$-$J_2$, Transverse-Field-Ising and $t$-$V$, and the homogeneous electron gas in two and three dimensions. The 411 lattice instances refer to different combinations of Hamiltonians, system sizes, lattice geometries (e.g.\ square, triangular, kagome), couplings, as well as boundary conditions (e.g.\ periodic, open). The 121 electron-gas instances hold 877 energies. The primary way of interacting with the leaderboard is through the webpage \url{https://www.qmbl.org}. Figures can be clicked and rows expanded to explore each result in more detail.

Each entry quotes the type of quantity (exact energy, variational bound, projected energy or extrapolation). The record-holding results are exact energies if they exist, or else the lowest energies (Fig.~\ref{fig:energies}); only energies flagged for a suspected error are skipped. Contributions of error bars can be made here~\cite{qmblContributing}. The rules are in \texttt{RULES.md}~\cite{qmblRules}.

\begin{figure*}[t]
  \centering
  \includegraphics[width=0.9\textwidth]{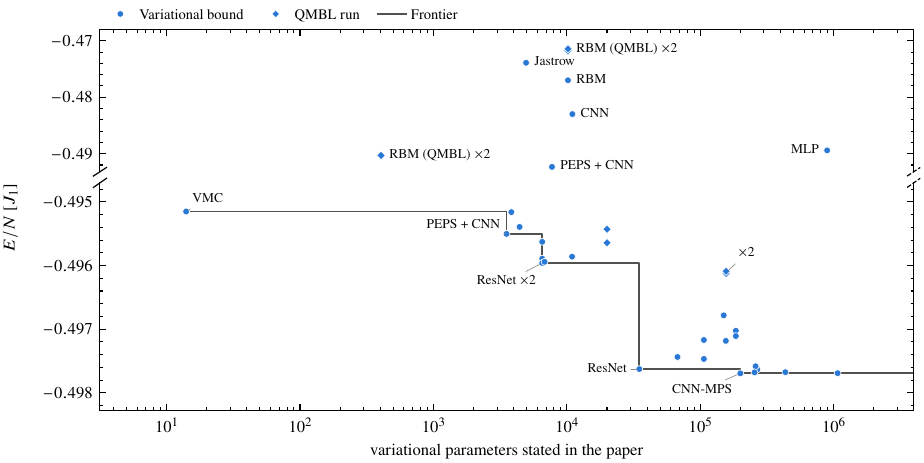}
  \caption{Exemplary figure of parameter counts of the same $10\times10$ $J_1$-$J_2$ lattice at $J_2=0.5$. The line is the frontier. ``(QMBL)'' marks QMBL's own estimation runs under one protocol, in case data was not available in papers. The interactive version also has the other two cost axes, hours and estimated FLOPs at \href{https://qmbl.org/\#energy-vs-compute}{qmbl.org/\#energy-vs-compute}.}
  \label{fig:params}
\end{figure*}

Estimations of compute cost were also collected, as well as calculated by us and can be found here~\cite{qmblCost} with the explanation of how we did it here~\cite{qmblFlops} (Fig.~\ref{fig:params}).

\par
\noindent\begin{minipage}{\columnwidth}
  \vspace*{14pt}
  \centering
  \includegraphics[width=\columnwidth]{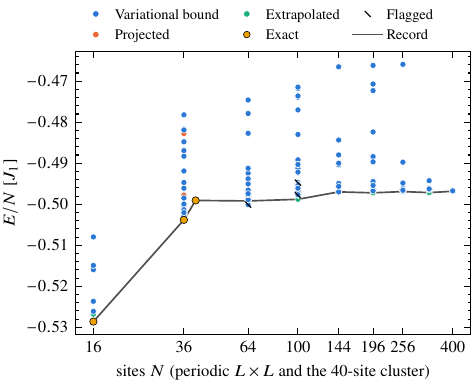}
  \captionof{figure}{As an example of the data you can find in QMBL, here is every published ground-state energy per site of the $J_1$-$J_2$ model at $J_2=0.5$ on periodic square lattices. Each dot is a paper's result. The records on each lattice are joined by the gray line. The interactive version with clickable dots is at \href{https://qmbl.org/\#energy-by-hamiltonian}{qmbl.org/\#energy-by-hamiltonian}.}
  \label{fig:energies}
\end{minipage}
\par\vspace{6pt}

The 285 instances without an exact energy are the targets of a companion benchmark, called QMBL: Open Problems, described in a separate paper. Everything is open source under this GitHub repository, including the results of agentic data sweeps~\cite{qmblRepo}. The data follow the FAIR principles for scientific data (findable, accessible, interoperable, reusable)~\cite{Wilkinson2016}, and, like the OptiMat Alloys database~\cite{Hu2026}, QMBL is a living database that records the provenance of every entry. Each release is archived on Zenodo with this paper describing v0.5.0~\cite{QMBL050}.

\onecolumngrid
\vspace{2ex}
\noindent
\begin{minipage}[t]{0.485\textwidth}
\begin{acknowledgments}
We thank Xiao Liang, Rong Cheng and Tao Li, P.\ S.\ Golubev, V.\ V.\ Mazurenko, and Javier Robledo Moreno for sending error bars, variances and measurement data. We encourage other authors to contribute through new results, error bars and corrections. Every energy in the table is the work of the groups cited on its row. We thank the authors of VarBench, whose compilation gave the starting set of references.
\end{acknowledgments}
\end{minipage}\hfill
\begin{minipage}[t]{0.485\textwidth}
\renewcommand{\refname}{References}
\renewcommand{\bibsection}{\section*{\refname}}
\raggedright

\end{minipage}

\end{document}